\documentclass[conference]{IEEEtran}

\usepackage{cite}
\usepackage{amsmath,amssymb,amsfonts}
\usepackage{amsthm}
\usepackage[ruled]{algorithm2e}
\usepackage{algpseudocode}
\usepackage{graphicx}
\usepackage{textcomp}
\usepackage{xcolor}
\usepackage{subcaption}
\usepackage{multirow}
\usepackage{booktabs}
\usepackage{makecell}
\usepackage{bbm}

\newcommand{\argmax}{\mathop{\mathrm{argmax}}}
\newcommand{\linebreakand}{%
  \end{@IEEEauthorhalign}
  \hfill\mbox{}\par
  \mbox{}\hfill\begin{@IEEEauthorhalign}
}

\begin{document}

\title{Discovering Command and Control Channels Using Reinforcement Learning}

\author{Cheng Wang$^{a}$$^{*}$,
        Akshay Kakkar$^{a}$,
        Chris Redino$^{a}$,
        Abdul Rahman$^{a}$,
        Ajinsyam S$^{a}$,        \\
        Ryan Clark$^{a}$,
        Daniel Radke$^{a}$,
        Tyler Cody$^{b}$,
        Lanxiao Huang$^{b}$,
        Edward Bowen$^{a}$\\
        \small $^{a}$Deloitte \& Touche LLP \\
        \small $^{b}$National Security Institute, Virginia Tech \\
        \small $^{*}$Corresponding author: chengwang@deloitte.com \\
}

\maketitle

\begin{abstract}

Command and control (C2) paths for issuing commands to malware are sometimes the only indicators of its existence within networks.  Identifying potential C2 channels is often a manually driven process that involves a deep understanding of cyber tradecraft. Efforts to improve discovery of these channels through using a reinforcement learning (RL) based approach that learns to automatically carry out C2 attack campaigns on large networks, where multiple defense layers are in place serves to drive efficiency for network operators. In this paper, we model C2 traffic flow as a three-stage process and formulate it as a Markov decision process (MDP) with the objective to maximize the number of valuable hosts whose data is exfiltrated. The approach also specifically models payload and defense mechanisms such as firewalls which is a novel contribution. The attack paths learned by the RL agent can in turn help the blue team identify high-priority vulnerabilities and develop improved defense strategies.
The method is evaluated on a large network with more than a thousand hosts and the results demonstrate that the agent can effectively learn attack paths while avoiding firewalls. 

\end{abstract}


\section{Introduction}

A command and control (C2) channel is made up of at least one path in a network that is designated for traffic flow consisting of commands from deployed malware to C2 server infrastructure \footnote{C2 infrastructure is made up of those servers that adversaries use to issue commands to deployed malware.}. As signature-less malware, \textit{i.e.} zero days, may evade detection by deployed countermeasures (e.g. endpoint detection and response, intrusion detection systems, intrusion prevention systems), the awareness of the C2 paths to deliver (operational) commands are sometimes the only indicators of malware within a network. Sophisticated cyber actors that operate as advanced persistent threats (APT)  carefully architect these C2 channels within networks to avoid detection for this reason and to perpetuate long term footholds within networks for a variety of purposes \cite{mitre-attack}.

This paper builds on previous work \cite{cody2022discovering, gangupantulu2021crown,route-surveillance, gangupantulu2021using} where reinforcement learning (RL) approaches were integrated with cyber terrain concepts \cite{conti_raymond_2018}. At the time of this writing, this work on C2 pathway discovery is known to be the first RL study to identify C2 channels within networks. The benefit of this work is to efficiently support and inform red (offense), blue (defense), and purple (integrated offense-defense) team operators on both offensive and defensive measures that could be implemented in support of discovering where possible C2 channels could exist. 

Instead of training a classifier to detect the presence of anomalous C2 traffic, this study attempts to examine \emph{how} C2 paths may be formulated within large networks using reinforcement learning. Identifying potential attack pathways on large networks is usually a very time consuming process when done manually. RL presents a viable option to automate this process through trial and error. By playing a red team role (\textit{i.e.}, offense), a trained RL agent can help discover vulnerable nodes in the network and suspicious activities, thereby helping the blue team (\textit{i.e.}, defense) develop plans to enhance, refine, or improve an organization's security posture.      

Our contributions are as follows. First, we develop a detailed reinforcement learning model that takes into account cyber defense terrain for multi-stage C2 attacks. Second, we demonstrate that our RL model can effectively identify attack paths on a large network,  which is generated using real-world data and has over 100 subnets and 1400 hosts. 

This paper is structured as follows. In Section \ref{sec:related} we review related work. Section \ref{sec:preliminaries} provides  background information on RL and C2.  In Section \ref{sec:methods}, we present the details of the proposed C2 simulation model and its RL formulation. Experiment results are discussed in Section \ref{sec:experiments}. Finally, we conclude the paper and discuss future work in Section \ref{sec:conclusion}.

\section{Related Work} \label{sec:related}
There has been a growing interest in applying RL to cyber security \cite{nguyen2019deep}. 
Compared with common machine learning models (e.g., Random Forest and Support Vector Machine, Naive Bayes), RL-based intrusion detection systems (IDS) \cite{sethi2020context, lopez2020application, alavizadeh2022deep} have shown better performance (in terms of accuracy, precision, recall, and F1 score)  on popular datasets such NSL-KDD \cite{tavallaee2009detailed} and the Aegean Wifi Intrusion Dataset (AWID) \cite{kolias2015intrusion}. Due to the classification nature of these models, the RL formulation is often straightforward: the state is the features (usually less than 50) selected from the given dataset, the action is the predicted class label, and the reward is positive (negative) for correct (incorrect) prediction.

RL has also been used for penetration testing. In particular, attack graphs \cite{mcdermott2001attack} and the Common Vulnerability Scoring System (CVSS) have emerged as an effective tool to construct the Markov Decision Process (MDP) for the RL agent \cite{yousefi2018reinforcement, chowdhary2020autonomous, hu2020automated, gangupantulu2021using}. To capture operational nuances, the notion of cyber terrain is first proposed in \cite{gangupantulu2021using}, where firewalls are treated as cyber obstacles and will incur protocol-specific negative rewards and reduce transition probabilities between states. Under this framework, successful applications have been made in crown jewels analysis \cite{gangupantulu2021crown} and identifying optimal paths for data exfiltration \cite{cody2022discovering}.

Cyber C2 has been a subject in the academic literature for decades \cite{vikelich2001architecture, erbacher2005extending}. Many works consider C2 simulation, control, and infrastructure \cite{bernier2012metrics, carvalho2013mtc2, carvalho2015mira}. More recent literature develops tools for automating command and control \cite{willett2015integrated, amro2022click}, including a non-proprietary language \cite{mavroeidis2020nonproprietary}. RL-based approaches to automating command and control have not been explored in the academic literature.

The literature on botnets often assumes a command and control setting \cite{shetu2019survey}, but botnets are not representative of all C2 operations. Moreover, botnets are often a tool to carry out command and control \cite{zeidanloo2009botnet}. In this way, in the context of this paper, botnets describe how to implement the more generically defined action space of the RL agent.

\section{Preliminaries} \label{sec:preliminaries}

\subsection{Reinforcement Learning}
Reinforcement learning is a framework in which an agent learns to optimize its behaviour by interacting with its environment \cite{sutton2018reinforcement}. The environment is usually modeled as a Markov decision process (MDP): $(\mathcal{S}, \mathcal{A}, P, r, \gamma)$, where $S$ is the state space, $\mathcal{A}$ is the action space, $P: \mathcal{S} \times \mathcal{A} \rightarrow \mathcal{S}$ is the transition function, $r: \mathcal{S} \times \mathcal{A} \times \mathcal{S} \rightarrow \mathbb{R}$ is the reward function and $\gamma \in (0,1]$ is the discount factor, which determines the present value of future rewards. The agent's behavior is characterized by its policy $\pi$, which is a probabilistic distribution over actions given a state. For deterministic policies, the action taken in state $s$ can be denoted as $\pi(s)$. At each time step, the agent observes a state $s_t$, takes an action $a_t$ according to $\pi(a|s_t)$, and transitions to a new state $s_{t+1}$ and receives a reward $r_t=r(s_t, a_t, s_{t+1})$. The cumulative discounted reward is called the \emph{return} and is defined as $G_t=\sum_{k=0}^\infty \gamma^k r_{t+k}$. The goal of an RL agent is to learn an optimal policy $\pi^*$ that maximizes the expected return from each state. Generally, RL algorithms can be categorized into three groups: value function-based (also known as critic-only) methods, policy gradient (or actor-only) methods, and actor-critic methods.

Value function-based methods such as $Q$-learning \cite{watkins1989learning} or deep Q-network (DQN) \cite{mnih2015human} learn optimal policies by first estimating the optimal action-value function $Q^*(s,a)$:
\begin{align}
    Q^*(s,a)& \equiv \max_\pi Q^\pi(s,a)  \nonumber \\
    &\equiv \max_\pi\mathbb{E}_\pi\big[G_t|s_t=s, a_t=a\big],
\end{align}
which can obtained by solving the Bellman equation:
\begin{align}
    Q^* (s,a) = \mathbb{E}_{s'} \big[ r + \gamma \max_{a'} Q^* (s',a') | s, a\big].
\end{align}
Then, an optimal policy $\pi^*$ is derived by selecting the action that yields the largest $Q$-value:
\begin{align}
    \pi^*(s) = \argmax_{a}Q^*(s,a).
\end{align}

Meanwhile, policy gradient approaches directly parameterize the policy $\pi(a|s;\theta)$ and optimize a performance measure $J(\theta)$ such as the expected return $\mathbb{E}[G_t]$ via gradient ascent. However, such methods often suffer from high variance and therefore may result in slow learning. To reduce the variance, actor-critic methods use an estimate of the value function $V_\pi(s)\equiv \mathbb{E}_\pi[G_t|s_t=s]$ as a baseline when estimating the policy gradient $\nabla J(\theta)$ \cite{nguyen2019deep}. 
The critic is responsible for learning the value function while the actor updates policy parameters by using the estimated policy gradient. In particular, the policy gradient can be estimated as 
\begin{align}
    \nabla J(\theta)  \approx \mathbb{E}\big[\nabla_{\theta}\log \pi(a_t|s_t; \theta)A_t\big],
\end{align} where $A_t=Q(s_t, a_t) - V(s_t)$ represents the \emph{advantage} of taking action $a_t$ at state $s_t$. 

One common problem with policy gradient methods is that large policy updates may occur and result in performance collapse, which can be difficult to recover from since the agent will be trained on the experience generated by bad policies. To improve training stability, Proximal Policy Optimization (PPO) \cite{schulman2017proximal} uses a clipped surrogate objective function:
\begin{align}
    \mathcal{L}(\theta) = \mathbb{E} \Big[ \min\big(\rho_t(\theta) A_t, \mathrm{clip}\big( \rho_t(\theta), 1-\epsilon, 1+ \epsilon \big) A_t\big)\Big], \label{eq:ppo_obj}
\end{align}
where $\rho_t(\theta) = \pi_\theta(a_t|s_t) / \pi_{\theta_{\mathrm{old}}} (a_t|s_t)$ is the probability ratio of the new policy over the old policy. The advantage function $A_t$  is often estimated using the generalized advantage estimation \cite{schulman2015high}, truncated after $T$ steps:
\begin{align}
    & \hat{A}_t = \delta_t + (\gamma\lambda) \delta_{t+1}+\cdots + (\gamma\lambda)^{T-t+1}\delta_{T-1},\\
    & \mathrm{where\;} \delta_t = r_t + \gamma V(s_{t+1}) - V(s_t).
\end{align}
To support exploration, an entropy bonus $\beta H(\theta)$ is often added to the objective function \eqref{eq:ppo_obj}, where $\beta$ is a coefficient.

\subsection{Command and Control} 
The term Command and Control (C2) has long been associated with military operations, referring to the systems in place and the use of the proper people overseeing the proper resources to accomplish specific mission goals \cite{stanton2008modelling}. C2, when applied to malware actors and behaviors, refers to the post-infection communication required for a piece of malware to work in concert with human orchestrators and effectively accomplish the goals. 
This goal may range from quietly sitting in place for a long period of time to a noisy smash-and-grab operation to get as much data as possible, regardless of the chance of getting caught. Dittrich and Dietrich \cite{dittrich2007command} examined the connection types and network traffic commonly associated with three distributed attacker/intruder tools (Handler/agent, central C2 mechanisms IIRC/botnet, P2P networks) and found that techniques that use direct, encrypted communications are the most difficult to locate and prevent, and a distributed external node C2 provides the most resiliency to having it’s communication lines severed.

To simulate real-world conditions in the internal networks, a series of communication paths, routers of various levels of security, and end-point nodes with potential vulnerabilities was created programmatically using a script. The conditions were created using real work experience from penetration testers and security analysts and adapted to accommodate the implementation of the 3-stage campaign model (detailed below). 
For this experiment the internal network nodes, once infected, require external communication to perform complex actions, but can still perform basic network enumeration and infection-spreading activities while waiting.

\section{Methods} \label{sec:methods}
In this section we present the details of C2 simulation model, its RL formulation, and the generation process of the test network. 

\subsection{Attack Simulation Overview}
\begin{figure}[t]
    \centering
    \includegraphics[width=.48\textwidth]{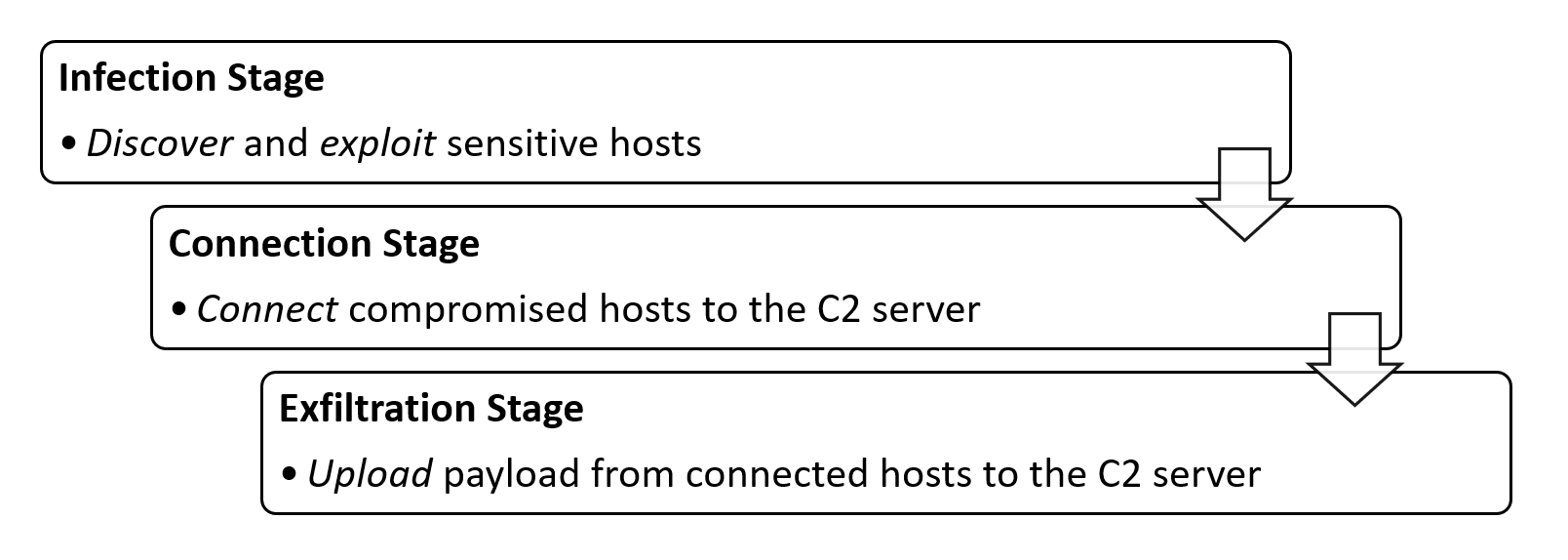}
    \caption{Command and control attack as a three-stage process.}
    \label{fig:c2overview}
\end{figure}

A C2 attack campaign is modeled as a three-stage process consisting of (i) infection, (ii) connection, and (iii) exfiltration (Fig. \ref{fig:c2overview}). The attacker first tries to gain a foothold on some target hosts by exploiting known vulnerabilities. It then seeks to establish communications with the C2 server for further instructions (e.g., lock or send out certain files). After identifying valuable information on the target system, the attacker starts to send data packets from the infected hosts to the remote server. An attack is considered to be successful if all three stages are completed in a given period of time. 

During the infection phase, the agent can perform a \emph{subnet scan} or an \emph{exploit} action on any given target. A subnet scan will reveal not only all hosts on the same subnet but also hosts with certain services on the adjacent subnets.
Each exploit is associated with a Common Vulnerabilities and Exposures (CVE) vulnerability and the success of the exploit depends on the presence of a specific service or process and the operating system running on the target host. A host must be discovered first through a subnet scan before it can be exploited. Likewise, doing a subnet scan from a particular host requires gaining access to it first. As a result, compromising a sensitive target often involves discovering and exploiting multiple intermediate hosts.

Once a sensitive host is compromised, the agent may initiate connection attempts by taking the \emph{connect} action, which sends a small packet to the C2 server on the Internet. To establish a connection, the traffic needs to pass through all firewalls between the host's subnet and the Internet. 
The connection attempt will be blocked if any one of the firewalls has gone through an update since the infection of the host. It may also fail with some probability, in which case another attempt is needed. However, too frequent connection attempts can raise alerts and lead to an emergency firewall update, which will not only block all future connections from the originating host but also from neighbouring infected hosts.

After establishing communications with the C2 server, a target payload of certain size is identified and a portion of the payload can be uploaded at a time. The task is complete when the entire payload is uploaded from every sensitive host. Similar to the connecting phase, outbound traffic are monitored by firewalls. Therefore, in order to send out the entire payload without being detected and blocked by firewalls, the agent needs to take deliberate pauses during the upload process. 

\subsection{Network Firewalls}
\begin{figure}[t]
    \centering
    \includegraphics[width=.48\textwidth]{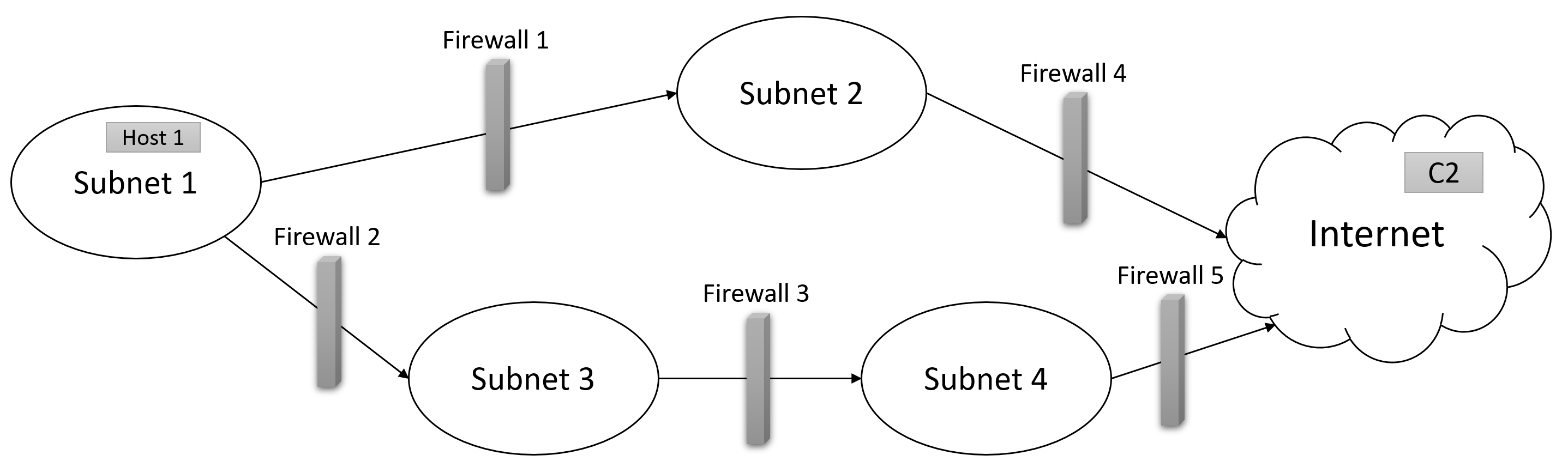}
    \caption{An example network with firewalls.}
    \label{fig:firewalls}
\end{figure}

Firewalls are located between all subnets and the Internet.
For the outbound traffic to reach to the C2 server, it needs to pass through \emph{all} firewalls on the shortest path (i.e., the path with the fewest hops) between the originating host's subnet and the Internet.
Fig. \ref{fig:firewalls} shows an example network with multiple firewalls, where traffic from Host 1 in Subnet 1 to the C2 server on the Internet needs to pass through both Firewall 1 and Firewall 4. 

Firewalls are updated periodically  or when unusual traffic patterns are detected. A wall-clock time (in seconds) is used to determine when a regular update is due. Unlike the number of time steps, which always increases by one in MDPs, the wall-clock time increases by different amounts depending on the actual action performed. 

An emergency firewall update will be triggered when one of the following conditions is met:
\begin{itemize}
    \item the cumulative connection attempts from a host exceeds a threshold \texttt{max\_connect\_attempts} (a connection attempt may be blocked by a firewall with probability $1 - \texttt{connect\_probability}$);
    \item the total upload volume during a five-minute window is greater than \texttt{max\_upload\_volume};
    \item the total upload time during a five-minute window is greater than \texttt{max\_upload\_time}.
\end{itemize}

The values of all firewall-related parameters are shown in Table \ref{tab:firewall}.
 \begin{table}[t]
     \centering
     \caption{Firewall Parameters.}
     \begin{tabular}{|c|r|} \hline
      \textbf{Firewall Parameter} &  \textbf{Value}\\ \hline
        connect\_probability & 0.8 \\
         max\_connect\_attempts & 3 \\
         max\_upload\_volume (MB) & 5000 \\
         max\_upload\_time (minutes) & 4 \\
         update\_frequency (hours) & 24\\ \hline
     \end{tabular}
     \label{tab:firewall}
 \end{table}
 
Both the scheduled and emergency firewall updates will patch security vulnerabilities, therefore blocking future connecting or uploading attempts from  compromised hosts. On the other hand, the attacker is assumed to be capable of adapting to the updates so that hosts compromised \emph{after} a firewall update may still establish connections to the C2 server. In short, there is only a limited window for the agent to connect to the C2 server after getting a foothold on a system.

\subsection{Reinforcement Learning Formulation}
\subsubsection{State Space}
The state includes the following features for every host: 
\begin{itemize}
    \item Subnet ID and local ID, 
    \item Operating system,
    \item Services and processes,
    \item Discovery value and status, 
    \item Infection value and status. 
\end{itemize}
Each host's subnet ID, local ID and operating system are one-hot encoded. Services and processes are represented by a vector of ones and zeros, where one means that the the service or process is running on the host. The discovery/infection status is changed from zero to one after the host is discovered/compromised. Similarly, discovery and infection values are the rewards given to the agent after it finds and exploits a host, respectively. 

In addition to the above features, each sensitive host has the following:
\begin{itemize}
    \item Connection status,
    \item Time since infection,
    \item Remaining payload size,
    \item Cumulative connecting attempts,
    \item Cumulative upload time and upload volume.
\end{itemize}
A sensitive host can be connected, not connected, or isolated. Its time since infection is measured in seconds instead of time steps. Data exfiltration is complete once the remaining payload size becomes zero. The cumulative metrics are computed with decay factor $d=0.999$. For example, let $c_{i,t}$ denote the cumulative connecting attempts at host $i$ and time step $t$, then 
\begin{align}
    c_{i, t+1}= 
    \begin{cases}
    c_{i,t} d^{\tau_{t, t+1}}+ 1, & \mathrm{if\;} a_{t+1} \mathrm{\;is\; to\;connect\; host\;}i\\
    c_{i,t} d^{\tau_{t, t+1}}, & \mathrm{otherwise}
    \end{cases}
\end{align}
where $\tau_{t, t+1}$ is the elapsed clock time between step $t$ and $t+1$.

\subsubsection{Action Space}
The action space consists of five types of actions: \emph{subnet scan} and \emph{exploit} actions during phase one of the attack campaign, \emph{connect} actions during phase two, \emph{upload} actions during phase three, and a \emph{sleep} action applicable in  all phases.
Except for the sleep action, which simply does nothing for a given period of time, each action requires specification of a target host. For a single host, there may be multiple exploit actions, one for each known vulnerability. Additionally, two upload actions are defined for each sensitive host - a fast one that uploads 1000MB of data in 10 seconds and a slow one that uploads 10MB in 10 seconds. 

As mentioned previously, after each step the simulation's clock time increases by a different amount based on the action performed. Table \ref{tab:actions} shows the wall-clock time for each action type considered in this study. Note that for erroneous actions, such as doing a subnet scan at a non-compromised host or uploading from a non-connected host, the simulation clock will only increase by one second as these errors will quickly interrupt the execution of the selected actions.   

\begin{table}[t]
    \centering
    \caption{List of actions.}
    \begin{tabular}{|c|c|r|} \hline
       \textbf{Action Type} & \textbf{Stage} & \textbf{Time} \\ \hline
       Subnet Scan & I& 30\\
       Exploit & I & 10 \\
       Connect & II & 1 \\
       Upload & III & 10\\
       Sleep & I, II, III & 60\\ \hline
    \end{tabular}
    \label{tab:actions}
\end{table}

\subsubsection{Reward Function}

The reward function can be broken into two parts: a reward for making progress towards the goal and a cost for taking a specific action. Actions with higher costs are more likely to trigger the defense system of the network. 

Gangupantulu et al. \cite{gangupantulu2021using} introduced cyber terrain into CVSS-based MDPs by modifying transition probabilities for traversing firewalls and the reward function for different protocols. Cody et al. \cite{cody2022discovering} further incorporated service-based defensive cyber terrain into CVSS-MDPs, which assumes that attackers can infer the presence of defenses based on the services running on a host even if they can not detect a defense directly. We adopt their methods and classify the services into three categories with high, medium, and low penalty. The cost of an action ranges from 1 to 6 and is determined by both its type (e.g, exploit or scan) and the services (e.g., http, imap, ssh) running on its target host.

The agent receives positive rewards after successfully discovering a target host, exploiting it,  connecting it, or uploading any partial payload from it to the C2 server. Upon finishing sending the entire payload, the agent is given a large bonus reward.  However, if exfiltration is detected by network firewalls, then the agent will receive a penalty equal to the total accumulated rewards gained on the originating host and the host will be isolated. That is, the agent will lose all rewards from discovery, infection, connection and partial uploads.
Table \ref{tab:rewards} lists rewards used in this study.

\begin{table}[t]
    \centering
    \caption{List of rewards.}
    \begin{tabular}{|c|r|} \hline
        \textbf{Reward Type} & \textbf{Value} \\ \hline
        Discovery &  1000\\ 
        Exploit & 1000\\ 
        Connection & 1000\\ 
        Upload (per unit) & 0.1 \\ 
        Upload (bonus) & 10000 \\ \hline
    \end{tabular}
    \label{tab:rewards}
\end{table}

\subsection{Network Generation}
To achieve the goal of generating realistic network topologies for C2 testing, 
we require networks to be realistic in size and scope, as well as in their composition, connectivity, and security posture.
To achieve these requirements, the following steps were taken during the initial network generation and configuration.

First, a series of variables were defined to control the size and scope of the network. Size and scope were measured by the following:
total (approximate) number of IPs, minimum and maximum number of IPs per subnet, and total number of subnets.
These variables would accept static values and the network generation script would use these values to composite a network that closely matched the desired size and scope. Randomization is essential to ensure that each network generated was unique.

Second, a series of variables were defined to meet the desired configuration of the network. These variables would accept static values to determine the maximum number of open ports and the maximum number of Common Platform Enumeration (CPE) types assigned to an IP address. A restriction is required so that the maximum number of designated CPE assignments will never be greater than the number of available open ports. Randomization is essential to ensure that each network generated was unique.

Third, a dataframe was constructed to hold the results of a continuous 24 hours data collection from the Shodan API. This dataframe was used to generate a reference dataframe that would be used to join CPEs by service and technology to each IP address. By applying a groupby function to the dataframe, the results were aggregated by port, service, and technology. The results of the groupby function were then used to construct the reference dataframe consisting of a record for each port number, a list of service \& technology combinations, and a probability for each item in the list. The reference dataframe was saved to a separate file for persistent use.

Fourth, a dataframe was constructed of ports and probabilities. The probability of each port's potential assignment to an IP address would be determined by the probability score of the port. The probability score of each port was determined by the open-frequency score in the ``nmap services" publication \cite{nampservices}.
Four buckets were defined to determine the probability of a port being assigned to an IP address:
High (0.1 - 1.0),
Moderate (0.05 - 0.1),
Low (0.005 - 0.05),
Rare (0.0 - 0.005).

To bring together these dataframes and variables into a cohesive network, a series of algorithms were defined. These were designed to implement the configuration of the created network and meet the desired connectivity of the network. The following is a list of the algorithms that were used to generate the network.
\begin{itemize}
    \item IP addresses were assigned to each subnet per the desired size and scope settings. These IP assignments would follow a general Class C netblock, defined as either a /24 CIDR notation or 255.255.255.0 netmask, to determine the correct IP address numbering scheme. The private addressing netblock 10.0.0.0/8 was used as the enterprise network to emulate common internal private IP usecases.
    \item The number of ports assigned to each IP address would be determined by the maximum number of open ports variable. A random number between 1 and the maximum number of open ports would be generated for each IP address. The random number would be used to determine the number of ports assigned to each IP address.
    \item To determine the assigned port a random number was generated between 0 and 1. This random generated number identifies the designated probability range for a given port assignment on an individual IP. A port number within the corresponding probability range would be assigned to the IP address. The algorithm would continue to generate random numbers until the maximum number of open ports was reached. The algorithm would then move on to the next IP address. This process would continue until all IP addresses were assigned the configured number of open ports.
    \item The dataframe of CPEs and port numbers was used to assign CPEs to each IP address' relevant open ports. The number of CPEs assigned to each IP address would be determined by the maximum number of CPEs variable. A random number between 1 and the maximum number of CPEs would be generated for each IP address. The random number would be used to determine the number of CPEs assigned to each IP address. If the number of CPEs assigned to an IP address was greater than the number of open ports, the number of CPEs would be reduced to match the number of open ports. The probability of each CPE being assigned to an IP address would be determined by the probability score of the CPE found within the corresponding reference dataframe.
    \item This CPE dataframe was further enriched by associating known security products to a new column in the dataframe. This column would be used to determine if a corresponding IP address was running a security product. A non-exhaustive list of security product descriptions identified includes: 
    firewall hardware and software, intrusion detection/prevention systems, and web application firewall software.
\end{itemize}

At this point in the network generation workflow, a network has been created consisting of a set number of subnets, numerous IP addresses assigned to each subnet, open ports assigned to each IP address, and CPEs assigned to the necessary open ports. In order to meet the desired connectivity of the network, an algorithm was created to determine the viability of service connections between different subnets. This algorithm assigned firewall allow rules to each subnet based on the desired connectivity settings. The following is a list of the connectivity settings that were used to determine the viability of service connections between different subnets:
\begin{itemize}
    \item Allow all traffic between subnets with mirrored services and technologies.
    \item Allow specific port traffic between subnets with individual matching services and technologies.
    \item Disallow traffic between subnets with no matching services and technologies.
\end{itemize}


To create vulnerabilities in the network, CVE exploits are assigned to each CPE in the network. The CVE exploits are attributed to each CPE by querying the CVE-Search database created locally on a server. CVE-Search \cite{cvesearch} is a tool to import CVE and CPE into a MongoDB to facilitate search and processing of CVEs.
The CVE exploits are parsed for the CVE ID, the CVSS score, and the CVSS vector. The CVE ID, CVSS score, and CVSS vector are then attributed in individual columns to the CPE record in a new dataframe. 



Once all facets of the network have been created, the network output is manifested in a singular yaml file. This file contains all of the information necessary to create testing scenarios. The yaml file contains the following information:
\begin{itemize}
    \item  Number of subnets,
    \item Number of IP addresses per subnet,
    \item Open ports per IP address,
    \item Associated services and technologies per open port,
    \item CPEs assigned to open ports,
    \item CVE exploits assigned to CPEs,
    \item CVSS scores assigned to CVE exploits,
    \item CVSS vectors assigned to CVE exploits,
    \item Firewall allow rules between subnets,
    \item Services and technologies designated as security products.
\end{itemize}

\section{Experiments} \label{sec:experiments}

In this section we describe the experiment details including the simulation network and the RL training procedure. We present the results across multiple random seeds and discuss key characteristics of the learned attack paths. 

\subsection{Network Description}

The experiment network has 101 subnets and a total of 1444 host. It is much larger than those used in previous studies, which usually have no more than a few hundred hosts. Each subnet contains  between 3 and 50 hosts. The attacker agent is assumed to have gained an initial foothold on host (1, 0) in subnet 1, which is directly connected to the Internet.
All other subnets are private and are not directly accessible from the Internet. 
A Windows machine (24, 3) from subnet 24 and a Linux machine (44, 5) from subnet 44 are selected as the C2 attack targets. Both machines are running HTTPS service and are not reachable from subnet 1.

\subsection{Training Details}
The RL agent is trained in an episodic fashion using the well-known PPO algorithm \cite{schulman2017proximal}. 
An episode ends when every sensitive host either completes sending payload to the C2 server or is isolated by firewalls. The target payload for each host is 10,000MB. To avoid extremely long episodes, a limit of 10,000 is imposed on the maximum number of time steps. 
Both the actor and the critic are approximated by a two-layer feed-forward neural network, where the first layer has 128 neurons and the second layer has 64 neurons. Other key hyperparameters are listed in Table \ref{tab:hyperparams}. The RL agent is trained for 5 million iterations and the training process is repeated 5 times with different random seeds.
All experiments are conducted on two Intel Xeon Platinum 8124M processors (18 cores/processor)\footnote{Comparable hardware could have been used for experiments}.   



\subsection{Results}
\begin{figure}[t]
    \centering
    \includegraphics[width=.48\textwidth]{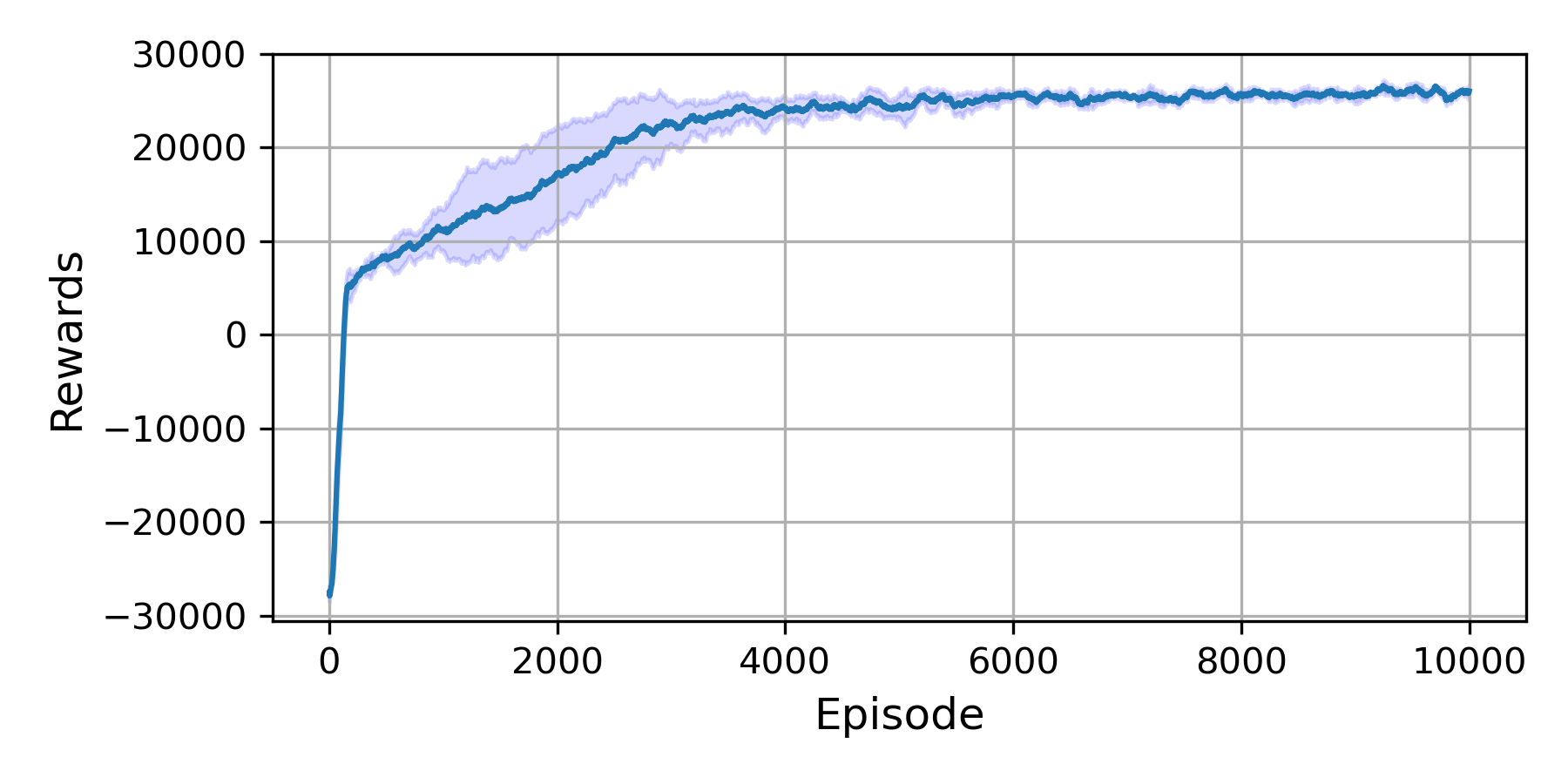}
    \caption{Average episode rewards over 5 runs.}
    \label{fig:rewards}
\end{figure}

\begin{figure}[t]
    \centering
    \includegraphics[width=.48\textwidth]{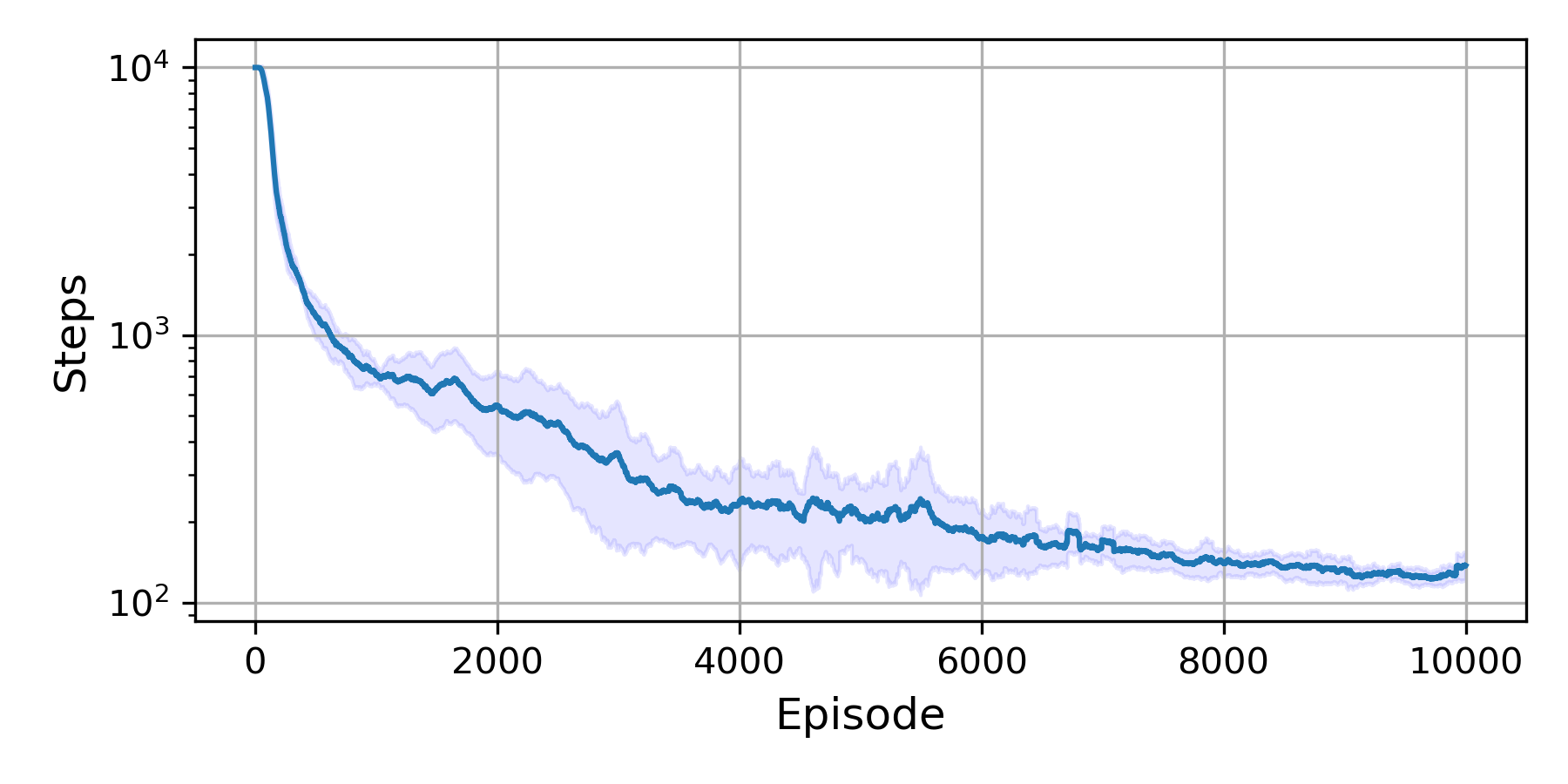}
    \caption{Average episode length over 5 runs.}
    \label{fig:steps}
\end{figure}

We report the average episode rewards over five training runs in Fig. \ref{fig:rewards} and the average episode length in Fig. \ref{fig:steps}. As can be seen, training is stable and the RL policy converges in 10,000 episodes. In particular, Fig. \ref{fig:rewards} shows that the sum of rewards in an episode steadily increases to 26,000, which is close to to the theoretical maximum under the reward structure listed in Table \ref{tab:rewards}. Meanwhile, the episode length gradually decreases, eventually averaging just over 100 steps (Fig. \ref{fig:steps}). This suggests that as training goes on, the RL agent completes the attack task  more efficiently and takes fewer random actions. 

\begin{table}[t]
    \centering
    \caption{List of hyperparameters.}
    \begin{tabular}{|l|l|} \hline
        \textbf{Hyperparameter} & \textbf{Value} \\ \hline
        Critic learning rate ($\alpha_w$) & $3\times10^{-4}$ \\ 
        Actor earning rate ($\alpha_\theta$) & $3\times10^{-5}$\\
        Discount factor ($\gamma$) & 0.99 \\
        Horizon (T) & 4096 \\ 
        Minibatch size & 64 \\
        Epochs (K) & 5 \\
        GAE parameter ($\lambda$) & 0.95 \\
        Clipping parameter ($\epsilon$) & 0.2 \\
        Entropy coefficient ($\beta$) & 0.001 \\ \hline
    \end{tabular}
    \label{tab:hyperparams}
\end{table}

To evaluate the final learned policy, we sample 100 attack paths using the trained actor network. Among these, 78 trajectories end with both target hosts completing sending payload to the C2 server, while in the other 22 scenarios one of the targets host is blocked by firewalls. In all cases, at least one target host is successfully attacked by the RL agent. Table \ref{tab:allpaths} reports statistics on the length, duration, and rewards from the generated attack paths. On average, the RL agent finishes the task in 107 steps or 47 minutes, and receives a total reward of 25,824.

\begin{table}[b]
    \centering
    \caption{Summary statistics of the generated attack paths.}
    \begin{tabular}{|l|r|r|r|} \hline
       & \textbf{Steps}  & \textbf{Duration (minutes)} & \textbf{Rewards} \\ \hline
      \textbf{Mean} & 107 & 47 & 25824 \\
      \textbf{Std} & 32  & 16 & 5005 \\ 
      \textbf{Min} & 69  & 27 & 15762 \\ 
       \textbf{Max} & 374  & 191 & 28859 \\ \hline
    \end{tabular}
    \label{tab:allpaths}
\end{table}

Due to the stochastic nature of the learned policy, the RL agent may take some unnecessary or redundant actions such as exploiting unimportant hosts or connecting a host that has already connected to the C2 server. After pruning the best-performing trajectory, we identify the key steps in the C2 attack as shown in Table \ref{tab:steps}. 

\begin{table}[b]
    \centering
    \caption{List of main steps taken by the RL agent.}
    \begin{tabular}{|l|c|l|} \hline
      \textbf{Action}  & \textbf{Target} &  \textbf{Vulnerability}  \\ \hline
         Subnet Scan &  (1, 0) & - \\
         Exploit & (92, 27) & CVE-2020-1259 \\
         Subnet Scan & (92, 27) & - \\
         Exploit & (44, 5) & CVE-2019-15920 \\
         Exploit & (24, 3) & CVE-2020-1259\\
         Connect & (44, 5) & - \\
         Connect & (24, 3) & - \\
         Upload (1000MB) & (44, 5) & - \\
         Upload (1000MB) & (24, 3) & - \\
         Sleep & - & - \\
     \hline
    \end{tabular}
    \label{tab:steps}
\end{table}

\begin{table}[t]
    \centering
    \caption{Statistics on the number of connect, upload, and sleep actions taken in successful episodes.}
    \begin{tabular}{|l|r|r|r|} \hline
     \textbf{Action}   & \textbf{Connect} & \textbf{Upload} & \textbf{Sleep} \\ \hline
      \textbf{Mean}  &  4.2 & 20  & 32.6   \\
      \textbf{Std}   & 1.6 & 0.0 & 1.5  \\ 
      \textbf{Min} & 2 & 20 & 29   \\ 
      \textbf{Max} & 10  & 20 & 36  \\ \hline
    \end{tabular}
    \label{tab:paths}
\end{table}

The agent starts the attack after getting an initial foothold on host (1,0) in subnet 1. It first performs a subnet scan, which leads to the discovery of hosts on other subnets. Among these, host (92, 27), a Windows machine, is selected as the target for further exploitation. After compromising this host, the agent does another subnet scan and finds both sensitive hosts (24, 3) and (44, 5), which are then exploited using different vulnerabilities. Once connections are established, the agent starts to upload payload from each host to the C2 server. Noticeably, the agent always take the fast upload option instead of the slow one. This is only possible if the agent knows when to sleep or pause to hide its activities.  

We summarize the number of connection attempts and the number of upload and sleep actions from  successful episodes in Table \ref{tab:paths}. On average, it takes 4.2 attempts before connections to the C2 server are established. 
Noticeably, the agent always chooses the fast option when it comes to uploading payload (20,000MB of payload for two targets requires at least 20 upload actions). Meanwhile, it takes many deliberate pauses, averaging over 32 sleep actions in an episode.

\begin{figure}[b]
    \centering
    \includegraphics[width=.46\textwidth]{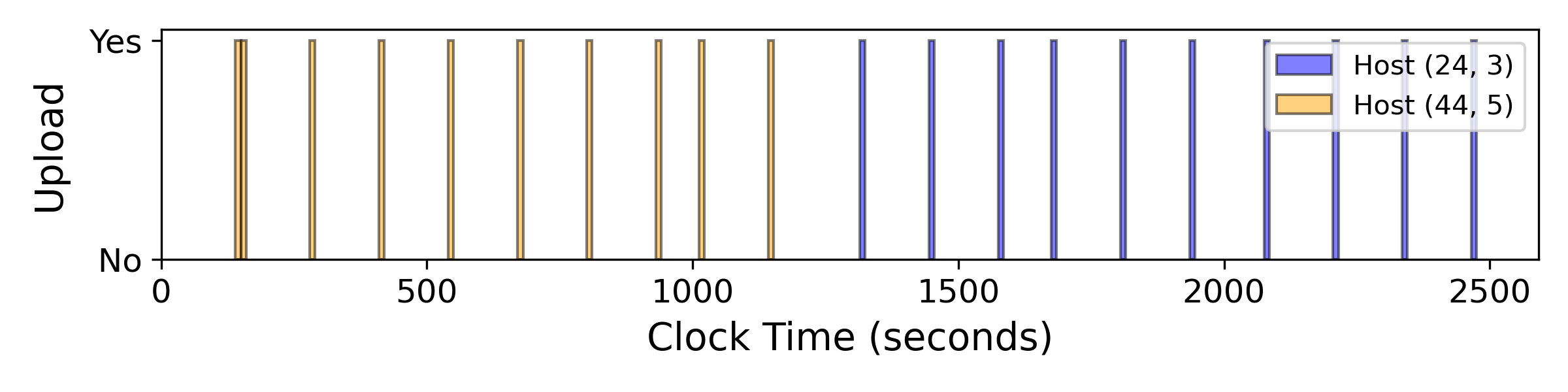}
    \caption{Times of upload actions taken during a C2 attack.}
    \label{fig:uploads}
\end{figure}

\begin{figure}[b]
    \centering
    \includegraphics[width=.46\textwidth]{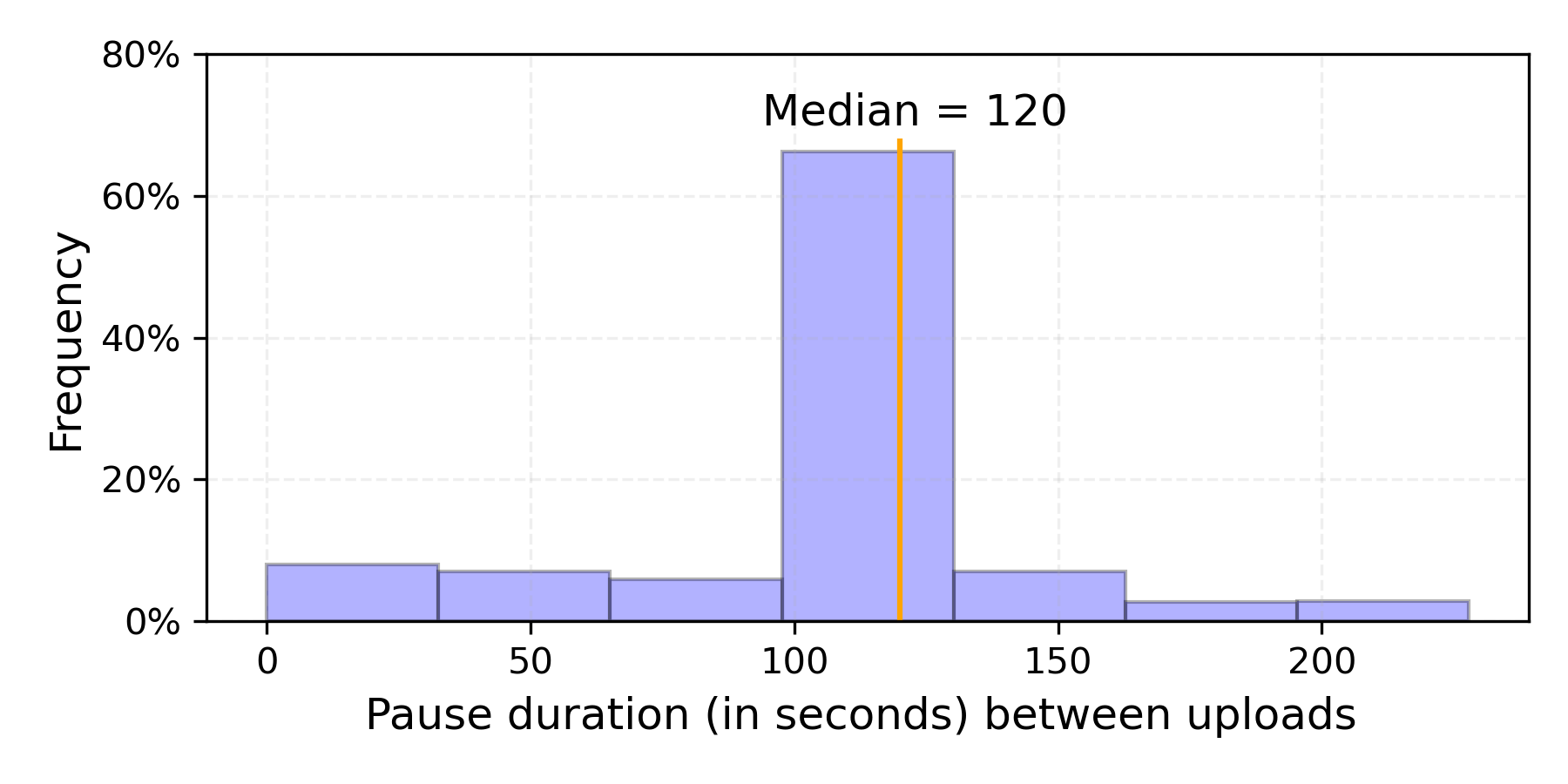}
    \caption{Distribution of the pause duration between upload actions from successful attack paths.}
    \label{fig:breaks}
\end{figure}

To further examine how the RL agent avoids firewall detection during uploads, we plot the time points at which upload actions are taken during a successful C2 attack in Fig. \ref{fig:uploads}. It is clear that the agent has learned to upload at a regular cadence to circumvent the current defense measures. The traffic pattern may then be further analyzed by security analysts to develop more sophisticated and effective defense strategies. For example, in addition to checking the traffic throughput during fixed time intervals, new firewall rules can be created based on the periodicity of traffic data.

The distribution of the time between consecutive uploads from  successful trajectories are shown in Fig. \ref{fig:breaks}. As we can see, most of the time the agent waits for about two minutes, which corresponds to taking two sleep actions, before resuming the upload process. This ensures that the agent is able to effectively use the available bandwidth for data exfiltration while keeping the total upload volume during the monitor window well below the alert threshold. 

\section{Conclusion} \label{sec:conclusion}
In this paper, we have proposed a reinforcement learning-based approach for discovering command and control attack paths. We showed that the RL agent can effectively complete the infection, connection, and data exfiltration stages in a large network without getting detected by firewalls. The identified attack path and traffic pattern can be further analyzed by security experts to discover new threats and develop enhanced security measures.

Future work should consider exfiltration using different protocols such as HTTPS and DNS. More sophisticated intrusion detection/prevention systems should also be incorporated into the simulation to provide a more realistic and challenging environment for the RL agent.

\bibliographystyle{IEEEtran}
\bibliography{ref}

\end{document}